\documentclass[aps,prc,twocolumn,amsmath,superscriptaddress,floatfix,nofootinbib]{revtex4-1}

\usepackage{setspace,ulem, epsfig,amssymb,amsfonts,amsmath,mathtools,bm,color,xcolor,graphicx,braket,adjustbox,esint,upgreek, verbatim,ctable}

\usepackage[colorlinks, linkcolor=red,anchorcolor=green,citecolor=blue]{hyperref}

\begin{document}

\title{Dynamical equation for quark spin polarization in the rotating medium}

\author{Tianyang Li}
\affiliation{Department of Physics, Tianjin University, Tianjin 300350, China}

\author{Yunfei Fan}\email{yunfei.fan@alumni.duke.edu}
\affiliation{Pratt School of Engineering, Duke University, Durham, NC 27708, USA}

\author{Anping Huang}\email{huanganping425@cumt.edu.cn}
\affiliation{School of Material Science and Physics, China University of Mining and Technology, Xuzhou 221116, China}

\author{Baoyi Chen}
\email{baoyi.chen@tju.edu.cn}
\affiliation{Department of Physics, Tianjin University, Tianjin 300350, China}

\date{\today}
\begin{abstract}
In non-central relativistic heavy-ion collisions, the produced quark-gluon plasma (QGP) behaves approximately as a rotating fluid due to the system's initial angular momentum. In this rotating fluid, the spins of quarks become polarized due to the coupling between spin and angular momentum, as well as random spin-spin interactions. Since the Landau-Lifshitz (LL) equation effectively describes the spin polarization of fermions in a medium with a magnetic field, we derive a phenomenological equation analogous to the LL equation for heavy quark spin dynamics in the rotating medium. The spin-angular momentum coupling and random spin-spin interactions are incorporated, leading to a detailed balance of heavy quark spin distributions. This equation provides insight into the spin dynamics of heavy quarks and quarkonium in relativistic heavy-ion collisions.
\end{abstract}

\maketitle

\section{Introduction}
Over the past few decades, extensive studies have demonstrated that a deconfined state of matter, composed of quarks and gluons—known as Quark-Gluon Plasma (QGP)—is produced in high-energy nuclear collisions~\cite{Bazavov:2011nk}. The transport properties~\cite{JET:2013cls}, equation of state of the hot QCD medium~\cite{Pang:2016vdc,Monnai:2019hkn}, and other related characteristics have been thoroughly investigated. The hot QCD medium is found to behave as a nearly perfect fluid, exhibiting almost zero viscosity, as evidenced by the large collective flows of light hadrons~\cite{ALICE:2011ab,CMS:2016fnw} and the substantial energy loss of heavy quarks~\cite{CMS:2017vhp,CMS:2018loe,ALICE:2018hbc}. In order to understand the yield and momentum distributions of light~\cite{Zhao:2017yhj,Ke:2018tsh} and heavy-flavor hadrons~\cite{Cao:2016gvr,He:2019vgs,Yang:2023rgb,Daddi-Hammou:2025hdz}, various theoretical studies have been conducted.

In non-central nuclear collisions, the hot deconfined medium also carries a large initial angular momentum~\cite{Liang:2004ph}. Hydrodynamic simulations predict that the vorticity of the rotating QGP can reach 10-20 MeV~\cite{jiang2016rotating}. This rotation of the QGP not only gives rise to a non-zero directed flow of hadrons in the forward and backward rapidity bins~\cite{STAR:2014clz}, but also induces a global spin polarization of quarks, attributed to the spin-angular momentum coupling. Both global~\cite{STAR:2017ckg} and local~\cite{CMS:2025nqr} spin polarization of the $\Lambda$ hyperon across different collision energies have been observed experimentally. Extensive studies have been conducted on the spin dynamics of light quarks and hadrons in heavy-ion collisions~\cite{Gao:2012ix,Huang:2018wdl,Liu:2019krs,Fu:2021pok,Sheng:2022wsy}.

Regarding heavy quarks and quarkonium, the rotation of the QGP also contributes to the direct flows of heavy quarks due to their strong coupling. This has been observed in  \(D\) mesons~\cite{STAR:2019clv}, suggesting that heavy quarks also carry angular momentum from the bulk medium despite their large mass. The spin of heavy quarks can be modified due to spin-angular momentum interactions. The partially spin-polarized heavy quarks can further hadronize into heavy quarkonium via the hadronization process, which can alter the spin alignment of vector mesons (such as \(J/\psi\)). Recently, the ALICE Collaboration measured the \(J/\psi\) spin alignment in Pb-Pb collisions~\cite{ALICE:2022dyy}. Previous theoretical studies with the assumption of the spin-polarized heavy quarks can explain the experimental data phenomenologically~\cite{Zhao:2023plc}. This highlights the need to develop a more effective phenomenological model for the detailed spin dynamics of heavy quarks and quarkonium in the hot QCD medium.

When considering the fermion spin dynamics in a magnetic field, the Landau-Lifshitz (LL) equation and the Landau-Lifshitz-Gilbert (LLG) equation serve as highly effective phenomenological models. These equations have proven successful in studies of magnetization and particle spin polarization, and are widely used in condensed matter physics~\cite{zhang2009generalization}. They have also been extended to describe heavy quark spin polarization in relativistic heavy-ion collisions~\cite{Liu:2024hii,Li:2025ipk,Dey:2025ail}. Concerning the spin polarization of fermions in a rotating medium, related studies have explored the form of the LL equation in a rotating reference frame~\cite{taniguchi2014magnetization,bertotti2001bifurcation}, which consider the effect of a rotating magnetic field from a classical perspective, but do not directly address the role of rotation in spin polarization. Therefore, we aim to develop an LL-like equation that can effectively describe the process of heavy quark spin polarization in the rotating medium.

This paper is organized as follows. In Section II, we first examine the Dirac equation in a rotating frame of reference and employ the Foldy-Wouthuysen (F-W) transformation to derive its non-relativistic Hamiltonian of heavy quarks. In Section III, we derive an LL-like equation applicable to rotating systems. In Section IV, the spin polarization of heavy quarks in a fermion gas with a fixed vorticity consisting of light quarks is numerically calculated. Finally, Section V provides a summary.  

\section{Theoretical model}


To examine heavy quark spin polarization in a rotating fermionic medium, an appropriate form of the Dirac equation is required in this rotating reference frame~\cite{chen2021charmonium,chen2016analogy},
\begin{equation}
ie_{a}^{\mu}\gamma^{a}\nabla_{\mu}\psi-m\psi=0,
\end{equation}
where the Latin index $a=(0,1,2,3)$ and Greek index $\mu=(t,x,y,z)$ represent the Lorentz indices in local flat space and curved space, respectively. \(\boldsymbol{\gamma}\) is the gamma matrix, and \(m\) is the mass of the fermion. \(e_{a}^{\mu}\) represants the tetrad field, and \(\nabla_{\mu}\) is the covariant derivative, given by \(\nabla_{\mu} = \partial_{\mu} + C_{\mu}\), where \(C_{\mu}\) is the affine connection. 

Using the duality relation, the specific form of \(e_a^\mu\) is as follows: \(e_{0}^{t} = 1\), \(e_{1}^{x} = 1\), \(e_{2}^{y} = 1\), \(e_{3}^{z} = 1\), \(e_{0}^{y} = -\omega x\), and \(e_{0}^{x} = \omega y\), where \(\omega\) is the angular momentum. The remaining terms in \(e_a^\mu\) are all zero. The corresponding form of the affine connection can be directly calculated as
\begin{equation}
C_{0}=-\frac{i}{2}\omega\sigma^{3},\quad C_{1}=C_{2}=C_{3}=0
\end{equation}
where $\sigma^3$ is one of the Pauli matrices. 
Finally, the Dirac equation in the rotational reference frame is simplified to be
\begin{equation}
i\gamma^\mu\partial_{\mu}\psi+\gamma^{0}\boldsymbol{\omega}\cdot\mathbf{\boldsymbol{s}}\psi-m\psi=0.
\end{equation}
And the Hamiltonian of the fermion in a rotating reference frame can also be written as
\begin{equation}
H=\boldsymbol{\alpha}\cdot\boldsymbol{p}+\beta m-\boldsymbol{\omega}\cdot\boldsymbol{j},
\end{equation}
with 
$\boldsymbol{j}=\boldsymbol{l}+\boldsymbol{s}$ to be the total angular momentum. \(\mathbf{l}\) and \(\mathbf{s}\) represent the angular momentum and spin of the fermion, respectively. \(\boldsymbol{\alpha}\) and \(\beta\) are the Dirac matrices. This derivation is consistent with the results in references~\cite{hehl1990inertial,ryder2008spin}. The Hamiltonian above also applies to the case of a time-dependent angular momentum \(\boldsymbol{\omega}(t)\).


Since the mass of the heavy quark is relatively large compared to the medium temperatures typically encountered in relativistic heavy-ion collisions, the non-relativistic form of the heavy quark Hamiltonian can be used when deriving the dynamical equations for heavy quark spin in the medium. The Foldy-Wouthuysen (F-W) transformation has been widely employed to obtain the non-relativistic form of the fermion Hamiltonian, while preserving the spin operator~\cite{nikitin1998exact,silenko2015general}. With the Foldy-Wouthuysen transformation, the Dirac equation is split into two simpler spinor equations. 
Finally, the non-relativistic form of the heavy quark Hamiltonian in the rotating medium is simplified to be,
\begin{equation}
H_{\rm NR}=\frac{\boldsymbol{p}^{2}}{2m}-\boldsymbol{\omega}\cdot\boldsymbol{j}.
\end{equation}
This Hamiltonian of the heavy quark will be used to derive the dynamical equations for the heavy quark spin in the following parts.

\section{Quantum LL-like equation }

The energy dissipation of heavy quarks in the medium can be described by a non-Hermitian Hamiltonian, \(\mathcal{H} = H_{\rm NR} - i\lambda \Gamma\), where \(\Gamma\) is a Hermitian operator and \(\lambda\) is a parameter. Similarly, the spin polarization process of heavy quarks in the rotating medium can be studied using an appropriate non-Hermitian form. However, since the normalization of the spin vector is conserved during the spin polarization process, the corresponding effective Hamiltonian for the heavy quark is constructed as,
$\mathcal{H}=H_{\rm NR}-i\lambda\left(H_{\rm NR}-\left\langle H_{\rm NR}\right\rangle \right)$ \cite{gisin1981spin,wieser2015description}. $\lambda$ corresponds to the rate of spin polarization along a certain direction. 

To describe the precession and polarization of the quark spin vector around the \(z\)-direction, which is chosen to be the direction of the medium's angular momentum, the unit vector representing the normalized spin vector \(\mathbf{S}\) is constructed as
\begin{equation}
{\bf S}/|{\bf S}|=\left(\sqrt{1-\eta^{2}} \cos \varphi ; \sqrt{1-\eta^{2}} \sin \varphi ; \eta\right)=\langle\psi| \boldsymbol{\sigma}|\psi\rangle
\end{equation}
where the time-dependent $\varphi(t)\in [0, 2\pi]$ describes the angle of the heavy quark precision process. The spin polarization, which corresponds to energy dissipation, is modeled by introducing time dependence in \(\eta(t) \in [-1, 1]\). In the case of complete spin polarization along the \(z\)-direction, \(\eta(t)\) becomes unity. \(\ket{\psi}\) is the spin part of the wave function of the fermion, and \(\boldsymbol{\sigma}\) represents the Pauli matrices.

\begin{figure}[!hbt]
    \centering
\includegraphics[width=0.4\textwidth]{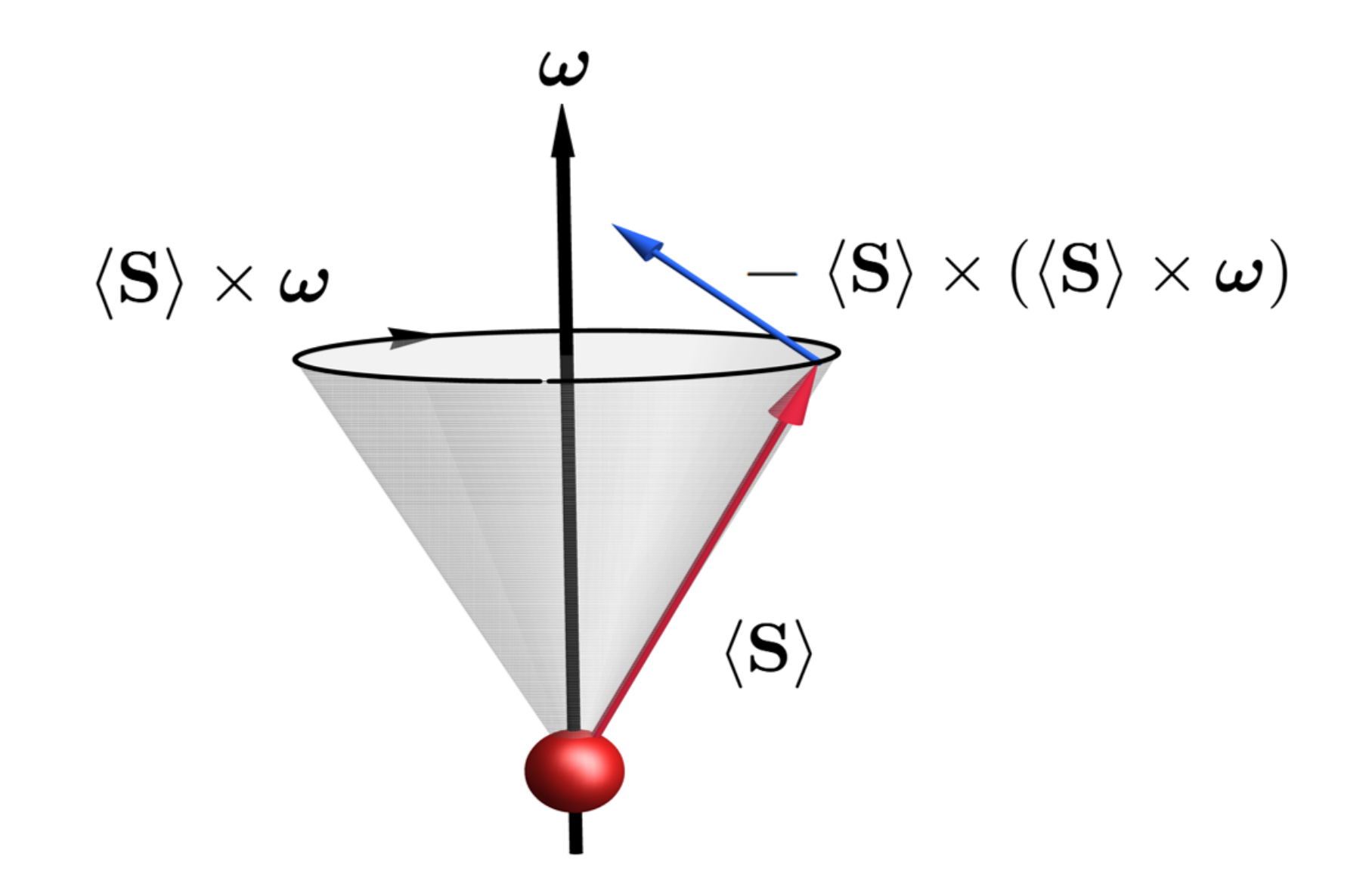}
    \caption{The precession and polarization of the heavy quark spin vector around the angular momentum \(\boldsymbol{\omega}\).}
    \label{lab-fig-2}
\end{figure}

The mean value of the heavy quark spin in the rotating fermion system is governed by the Hamiltonian,
\begin{align}
    \frac{d}{d t}\langle{\mathbf{S}}\rangle&=\operatorname{Tr}\left(\frac{d}{d t} {\rho} {\mathbf{S}}\right)\nonumber\\
    \label{lab-eq-st}
&=i\operatorname{Tr}\left([\rho,H_{\rm NR}]\mathbf{S}\right)-\lambda\operatorname{Tr}\left([\rho,[\rho,H_{\rm NR}]]\mathbf{S}\right).
\end{align}
where $\rho$ is the spin density operator. The second line in the above equation employs the Liouville equation, 
\begin{align}
\frac{d\rho}{dt} & =\frac{d}{dt}(|\psi\rangle\langle\psi|)=\frac{d|\psi\rangle}{dt}\langle\psi|+|\psi\rangle\frac{d\langle\psi|}{dt}\nonumber \\
 & =i[\rho,H_{\rm NR}]-\lambda[\rho,[\rho,H_{\rm NR}]].
 \label{drou/dt}
\end{align}
To calculate the spin density operator \(\rho\) of the fermionic system and substitute it back into Eq.~(\ref{lab-eq-st}), we follow Ref.~\cite{hofmann2004quantum} to perform the expansion of \(\rho\) for the fermionic system,
\begin{equation}
    {\rho}=\frac{1}{2 S+1} {\mathbf{1}}+\frac{1}{n_{S}} \sum_{i}\left\langle{S}_{i}\right\rangle {S}_{i}+\frac{1}{n_{2 S}} \sum_{ij}\left\langle{S}_{ij}\right\rangle {S}_{i j}+\ldots
\end{equation}
 with the definition
 \begin{equation}
  {S}_{ij}=\frac{1}{2}\left[{S}_{i}, {S}_{j}\right]_{+}-\frac{1}{3} S(S+1) \delta_{ij}.   
 \end{equation}
Here $ i, j \in\{x, y , z\}$ and $\left[{S}_{i}, {S}_{j}\right]_{+}={S}_iS_j+S_jS_i$. $n_S$ and $n_{2S}$ are defined as $n_{S}\equiv \operatorname{Tr}\left({S}_{i} {S}_{i}\right)$ and $n_{2 S}\equiv \operatorname{Tr}\left({S}_{ij} {S}_{ij}\right)$. In the system with $S=1/2$, $S_{ij}\equiv 0$, and its density operator becomes
\begin{equation}
\rho=\frac{1}{2}\left(\mathbf{1}+\sum_{i}\left\langle \sigma_{i}\right\rangle \sigma_{i}\right),
\end{equation}
Substituting the simplified density operator into \(\frac{d\langle \mathbf{S} \rangle}{dt}\) and retaining the spin part of the Hamiltonian \(\mathcal{H}\), the time evolution of the heavy quark spin vector becomes,
\begin{align}
\label{lab-eq-ll}
\frac{d\left\langle \mathbf{\boldsymbol{S}}\right\rangle }{dt}=\left\langle\mathbf{\boldsymbol{S}}\right\rangle \times\boldsymbol{\omega}-\lambda\left\langle \mathbf{\boldsymbol{S}}\right\rangle \times(\left\langle \mathbf{\boldsymbol{S}}\right\rangle \times\boldsymbol{\omega}).
\end{align}
The first term describes the spin precession around the direction of the medium's angular momentum \(\boldsymbol{\omega}\), while the second term describes the quark spin polarization along the direction of \(\boldsymbol{\omega}\). This is illustrated in Fig.~\ref{lab-fig-2}. \(\lambda\) can be interpreted as the spin-damping factor.

This equation, Eq.~(\ref{lab-eq-ll}), closely resembles the Landau-Lifshitz (LL) equation describing the fermion spin dynamics in a fermion system with an external magnetic field~\cite{gaididei1999noise,grinstein2003coarse}. Since fermion-fermion interactions are absent in the Hamiltonian we constructed, the noise term—responsible for reducing spin polarization due to random spin-spin interactions—does not appear in Eq.~(\ref{lab-eq-ll}). The values of the damping factor \(\lambda\), which characterizes the rate of spin polarization, have been preliminarily explored in previous work~\cite{Li:2025ipk}. Here, we directly incorporate the noise term into Eq.~(\ref{lab-eq-ll}), analogously to the LLG equation in the magnetic field, and derive a phenomenological equation for the heavy quark spin dynamics in the rotating fermion system,
\begin{align}
\frac{d\left\langle \mathbf{\boldsymbol{S}}\right\rangle }{dt}=\left\langle \mathbf{\boldsymbol{S}}\right\rangle \times\left(\boldsymbol{\omega}+\boldsymbol{\mathrm{\omega}}_{\mathrm{th}}\right)-\lambda\left\langle \mathbf{\boldsymbol{S}}\right\rangle \times(\left\langle \mathbf{\boldsymbol{S}}\right\rangle \times\left(\boldsymbol{\omega}+\boldsymbol{\mathrm{\omega}}_{\mathrm{th}}\right)).
\label{LL-likenoise}
\end{align}
The three-dimensional term \(\boldsymbol{\omega}_{\mathrm{th}}\) is assumed to be white noise that follows a Gaussian distribution, and its strength is determined by the fluctuation-dissipation relation~\cite {nishino2015realization},
\begin{align}
\label{avomega}
\left\langle \omega_{{\rm th},\mu}(t)\right\rangle &=0,\\
\label{avsqomega}
\left\langle \omega_{{\rm th},\mu}(t)\omega_{{\rm th},\nu}\left(t^{\prime}\right)\right\rangle &=C_{Q}{2\lambda k_{B}T}\delta_{\mu\nu}\delta\left(t-t^{\prime}\right),
\end{align}
where $\mu,\nu$ is the index representing three dimensions. 
\(k_B\) is the Boltzmann constant, \(T\) is the temperature of the system, and \(t, t'\) are the time points. In the long-time limit, the spins of the quarks follow the distribution \(\exp\left(-\beta \mathcal{H}\right)\), where \(\beta = \frac{1}{k_B T}\), and \(\mathcal{H}\) is the Hamiltonian of heavy quarks located in the vortical and noise fields, represented by \(\boldsymbol{\omega}\) and \(\boldsymbol{\omega}_{\mathrm{th}}\), respectively. This equilibrium distribution determines the strength of the noise terms.

The above fluctuation-dissipation relation and the Gaussian-like noise are suitable for classical variables, such as particle momentum which is continuous. However, for the case of quark spin along the \(z\)-direction, the mean value of the quark spin \(\langle S_z \rangle \equiv \frac{\sum_i S_z^i \exp(-\beta \mathcal{H}_i)}{\sum_i \exp(-\beta \mathcal{H}_i)}\) differs depending on whether a continuous or discrete spins are employed in the calculation. The ratio of the \(\langle S_z \rangle\) values with quantum and semiclassical \(S_z\) distributions for fermions with spin-1/2 is calculated to be,
\begin{align}
    \frac{\langle S_z\rangle_{\rm classical}}{\langle S_z\rangle_{\rm quantum}}=\frac{1}{3}
\end{align}
More details are contained in the Appendix. \ref{appendix-a}. To mimic the quantum effect associated with the discrete spin distribution, we introduce the factor \(C_Q = 1/3\) in the noise term, which enhances the value of \(\langle S_z \rangle\). With the combined effects of the polarization term \(\lambda\) and the noise term \(\boldsymbol{\omega}_{\mathrm{th}}\), the spin of the quarks in the rotating medium approaches equilibrium with detailed balance.

\section{Numerical results of LL-like equation} 

In this section, we employ an LL-like equation to numerically evolve the spin of heavy quarks in the rotating medium. Instead of using realistic temperature evolutions for the medium, as given by hydrodynamic models, we study the case of a static medium with a constant temperature. At RHIC collision energies, the typical temperature of the bulk medium is around \(T = 0.2 \, \text{GeV}\) in semi-central nuclear collisions. The temperature is encoded in the strength of the noise term in the LL-like equation, which reduces the degree of quark spin polarization along the direction of the medium’s angular momentum.

Regarding the medium's angular momentum \(\omega\), extensive studies have been conducted to extract its value across different collision energies. For this preliminary study, we assume \(\omega = 20 \, \text{MeV}\) as a constant and neglect its time and spatial dependence. Another important factor affecting the spin dynamics of heavy quarks is the polarization rate \(\lambda\). Since there is no reference for calculating the values of \(\lambda\) for heavy quarks in hot quark matter controlled by the strong force, we explore different values of the damping factor, \(\lambda = (0.1, 0.05, 0.03)\), in the following numerical simulations. The effect of quark mass on the spin dynamics is already incorporated into \(\lambda\). Note that \(\lambda\) not only determines the rate of quark spin polarization but also alters the strength of the noise term.

\begin{figure}[!htb]
\includegraphics[width=0.42\textwidth]{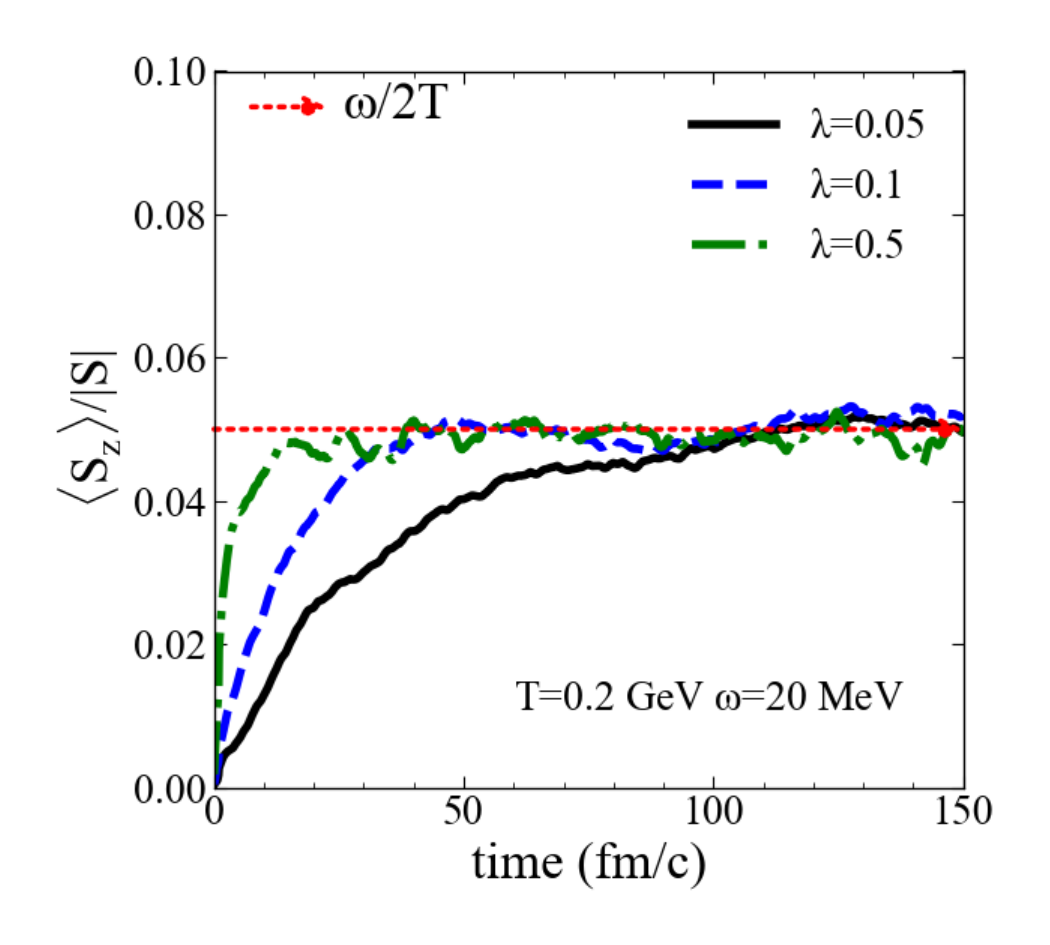}
\caption{ 
The time evolution of the mean value of the normalized heavy quark spin vector \(\langle S_z \rangle / |{\bf S}|\) along the \(z\)-direction in rotating quark matter with a constant temperature \(T = 0.2 \, \text{GeV}\) and a constant angular momentum \(\omega = 20 \, \text{MeV}\), with the angular momentum along the \(z\)-direction. The damping factor is taken as \(\lambda = 0.03, 0.05, 0.1\), respectively.
}
\label{lab-s-t}
\end{figure}

In Fig.~\ref{lab-s-t}, we evolve the spin of quarks in a rotating medium with angular momentum \(\omega = 20 \, \text{MeV}\). \(\langle S_z \rangle / |S|\) represents the normalized value of the quark spin along the \(z\)-direction, averaged over a large number of events (one million quark events are simulated). The three lines correspond to three values of the damping factor. The fluctuations in the lines are induced by the noise terms in the LL-like equation, which arise from random spin-spin interactions between fermions. As shown in Fig.~\ref{lab-s-t}, with increasing damping factor, the heavy quark spin becomes polarized more rapidly. However, in the long-time limit, the value of \(\langle S_z \rangle / |S|\), which represents the degree of spin polarization in the equilibrium stage, approaches a similar value across all three lines. This value is also consistent with the relation \(\sim \omega / (2T)\) given by equilibrium theories~\cite{becattini2020polarization}, which is marked by thin dashed line in the figure.

  \begin{figure}[!htb]
\includegraphics[width=0.42\textwidth]{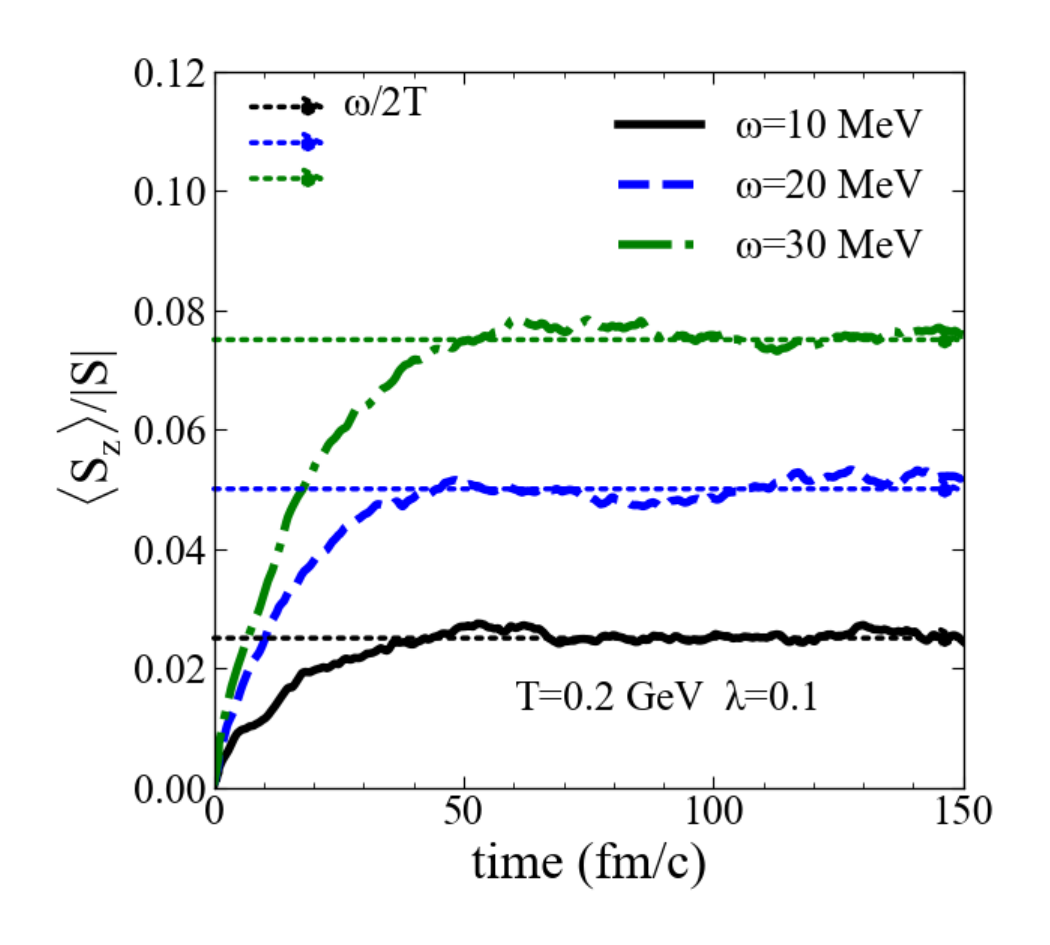}
\caption{ 
The spin polarization of quarks plotted over time, with the parameters $\lambda=0.1$, $T=0.2$ GeV, and $\omega=10,20,30$ MeV. 
}
\label{lab-wvar}
\end{figure}

As the global angular momentum of the medium generated in heavy-ion collisions varies significantly with collision energy, we consider different values of the angular momentum \(\omega = (10, 20, 30) \, \text{MeV}\) in Fig.~\ref{lab-wvar}. With larger values of \(\omega\), the degree of heavy quark spin polarization increases. This trend is consistent with the expression \(\omega / (2T)\), which are plotted and compared in the figure.

Note that although the polarization rate \(\lambda\) does not affect the final degree of spin polarization in the equilibrium limit of the static medium, in the case of heavy-ion collisions—where the medium temperature decreases rapidly with time and the angular momentum of the medium varies with coordinates—different values of \(\lambda\) can lead to different final spin distributions of heavy quarks after they exit the medium. Furthermore, unlike the magnetic field, which results in opposite spin polarizations for quarks and antiquarks, the spin-angular momentum coupling leads to similar spin polarizations for both quarks and antiquarks. This coupling can therefore alter the spin alignment of \(J/\psi\) via the charm quark coalescence process, which has been measured in experiments. Consequently, our numerical simulations provide valuable insights into the understanding of \(J/\psi\) spin alignment in heavy-ion collisions.

\section{Summary}

In summary, the hot deconfined medium produced in semi-central nuclear collisions carries a substantial angular momentum. The spin of quarks can become partially polarized in the rotating QCD medium due to spin–angular momentum coupling and spin–spin interactions between the quarks. To phenomenologically study this effect, a non-Hermitian Hamiltonian for heavy quarks is introduced to describe their spin dynamics in the rotating medium. The corresponding equation, analogous to the Landau-Lifshitz (LL) equation in a magnetic field, is presented. In a static medium with constant temperature and angular momentum, although the process of heavy quark spin polarization is closely related to the damping factor in the LL-like equation, the magnitude of the heavy quark spin polarization in the equilibrium limit is proportional to the ratio $\omega/T$. Furthermore, the magnitude of the quark spin polarization is examined for different values of angular momentum. This research provides valuable insights for future studies of heavy quark spin polarization and charmonium spin alignment in relativistic heavy-ion collisions.

 \vspace{0.5cm}
 {\bf Acknowledgement:}
 This work is supported by the National Natural Science Foundation of China (NSFC) under Grant No. 12175165 and 12575149.

\vspace{0.5cm}

\appendix
\section{The noise term in the stochastic LL-like equation}
\label{appendix-a}

The strength of the noise term in the LL-like equation can be calculated with the fluctuation-dissipation theorem\cite{nishino2015realization}.
We know that the noise term will have the following form,
\begin{align}
\left\langle \omega_{th,\mu}(t)\right\rangle &=0,\nonumber \\
\left\langle \omega_{th,\mu}(t)\omega_{th,\nu}\left(t^{\prime}\right)\right\rangle &=2\kappa\delta_{\mu\nu}\delta\left(t-t^{\prime}\right),
\end{align}
where the strength of the noise term $\kappa$ will be determined in the following part.
The spin of the $i$-th fermion in the system satisfies the equation,
\begin{equation}
\frac{ds_{i}^{\mu}}{dt}=f_{i}^{\mu}\left(\boldsymbol{s}_{i}\right)+g_{i}^{\mu v}\left(\boldsymbol{s}_{i}\right)\omega_{th,v;i}(t),
\label{a9}
\end{equation}
where $f_{i}^{\mu}\left(\boldsymbol{s}_{i}\right)=\left(\mathbf{s}_i\times\boldsymbol\omega-\lambda|S|\mathbf{s}_i\times (\mathbf{s}_i\times\boldsymbol\omega )\right)^{\mu},g_{i}^{\mu v}\left(\boldsymbol{s}_{i}\right)=\epsilon^{\mu\rho\nu}s_{i\rho}+\lambda\delta^{\mu\nu}-\lambda s_i^\mu s_i^\nu,\mathbf{s}=\mathbf{S}/{|S|}.$
In the equation, \(\mu,\nu =1,2,3\) for x, y, z directions. $\epsilon^{\mu\rho\nu}$ is the Levi-Civita symbol.
It is easy to see that this form is very similar to the Langevin equation in fluid dynamics. Now, we consider the following form of the probability distribution function $F=F(\boldsymbol{s_1},\boldsymbol{s_2},...\boldsymbol{s_N},t)$, which satisfies the continuity equation of the distribution: 
\begin{equation}
\frac{\partial}{\partial t}F+\sum_{i=1}^{N}\frac{\partial}{\partial s_{i}^{\mu}}\left\{ \left(\frac{d}{dt}s_{i}^{\mu}\right)F\right\} =0.
\end{equation}
 After substituting into the LL-like equation, it can also be written as: 
\begin{equation}
\frac{\partial}{\partial t}F+\sum_{i=1}^{N}\frac{\partial}{\partial s_{i}^{\mu}}\left\{ \left(f_{i}^{\mu}\left(\boldsymbol{s}_{i}\right)+g_{i}^{\mu v}\left(\boldsymbol{s}_{i}\right)\omega_{th,v;i}(t)\right)F\right\} =0.
\label{Fform}
\end{equation}

The statistical average of the probability distribution $P=\left\langle F\right\rangle$ and its continuity equation becomes
\begin{equation}
\frac{\partial}{\partial t}P=-\left\langle \sum_{i=1}^{N}\frac{\partial}{\partial s_{i}^{\mu}}\left\{ \left(f_{i}^{\mu}\left(\boldsymbol{s}_{i}\right)+g_{i}^{\mu v}\left(\boldsymbol{s}_{i}\right)\omega_{th,\nu,i}(t)\right)F\right\} \right\rangle 
\end{equation}
Based on Eq. \ref{a9}, we have the corresponding Fokker-Planck equation:
\begin{equation}
\frac{\partial}{\partial t}P=-\sum_{i=1}^{N}\frac{\partial}{\partial s_{i}^{\mu}}\left\{ f_{i}^{\mu}P-\kappa g_{i}^{\mu\nu}\frac{\partial}{\partial s_{i}^{\sigma}}\left(g_{i}^{\sigma\nu}P\right)\right\} 
\label{Pform}
\end{equation}
According to the definition of $g_{i}^{\mu v}$, we have $g_{i}^{\mu\nu}\frac{\partial}{\partial s_{i}^{\sigma}}g_{i}^{\sigma\nu}=0.$
Finally, we have: 
\begin{equation}
\begin{aligned}\frac{\partial}{\partial t}  P
= & -\sum_{i}\frac{\partial}{\partial\boldsymbol{s}_{i}}  \left\{ \left[\boldsymbol{s}_{i}\times\boldsymbol{\omega}-\lambda|S|\mathbf{s}_i\times (\mathbf{s}_i\times\boldsymbol\omega)\right.\right.\\
 &  +\kappa\boldsymbol{s}_{i}\times (\boldsymbol{s}_{i}\times\frac{\partial}{\partial\boldsymbol{s}_{i}} ) ]P \} 
\end{aligned}
\label{eq:llg-p1}
\end{equation}
Next, we consider that after a long time, the system reaches equilibrium, which gives $P\propto\exp\left(-\beta\mathcal{H}\right)$. The sum of the second and third terms in Eq.\ref{eq:llg-p1} is 0. Then, 
\begin{equation}
\frac{\partial}{\partial\boldsymbol{s}_{i}}P\left(\boldsymbol{s}_{1},\ldots,\boldsymbol{s}_{N}\right)=\beta|S|\boldsymbol{\omega}P\left(\boldsymbol{s}_{1},\ldots,\boldsymbol{s}_{N}\right).\label{eq:llg-p}
\end{equation}
substitute Eq.\ref{eq:llg-p} into Eq.\ref{eq:llg-p1}, one determine the strength of the noise term
\begin{equation}
\kappa={\lambda k_BT}.
\end{equation}

In the calculation of the mean value of the spin $\langle S_z\rangle$, when we use the semiclassical approximation of spin, under the weak field approximation, the mean value can be calculated to be, 
\begin{equation}
    \begin{aligned}
\left\langle S_{z}\right\rangle_{\rm classical} & =\frac{\iint|S| \cos \theta e^{\frac{\omega|S| \cos \theta}{T}} \sin \theta d \theta d \phi}{\iint e^{\frac{\omega|S| \cos \theta}{T}} \sin \theta d \theta d \phi} \\
& \approx \frac{\omega}{12 T}
    \end{aligned}
\end{equation}
In the quantum case where $S_z$ becomes discrete, the value of $\langle S_z\rangle$ becomes
\begin{equation}
    \begin{aligned}
\left\langle S_{z}\right\rangle_{\rm quantum} & =\frac{\Sigma_{m=-1 / 2,1 / 2} e^{\frac{\omega m}{T}} m}{\Sigma_{m=-1 / 2,1 / 2} e^{\frac{\omega m}{T}}} \\
& \approx \frac{\omega}{4 T} .
\end{aligned}
\end{equation}

To account this quantum effect in the distribution of $S_z$, we introduce a factor $\langle S_z\rangle_{\rm classical}/\langle S_z\rangle_{\rm quantum}$ in the noise term, 
\begin{equation}
\kappa=\frac{\lambda k_BT}{3}.
\end{equation}

Further studies will be performed by incorporating quantum noise.

\bibliography{sub1-vortical}



\end{document}